\documentclass[preprint,showpacs,preprintnumbers,amsmath,amssymb]{revtex4} % newly added

\usepackage{CJK}
\usepackage{graphicx}% Include figure files
\usepackage{dcolumn}% Align table columns on decimal point
\usepackage{bm}% bold math
\usepackage{ifthen}
\usepackage{booktabs}

\begin{document}

\title{Spin dynamics in the XY model}
\author{Rong-Chun Ge, Chuan-Feng Li$\footnote{email: cfli@ustc.edu.cn}$, Guang-Can Guo}
\affiliation {Key Laboratory of Quantum Information, University of
Science and Technology of China, CAS, Hefei, 230026, People's
Republic of China}
\date{\today }
%\pacs{68.65.Hb, 73.22.-f, 78.67.Hc }% PACS, the Physics and Astronomy
% 68.65.-k Low-dimensional, mesoscopic, and nanoscale systems:
%          structure and nonelectronic properties
% 68.65.Hb Quantum dots
% 73.22.-f Electronic structure of nanoscale materials: clusters,
%          nanoparticles, nanotubes, and nanocrystals
% 73.21.La Electron states and collective excitations in multilayers, quantum
%          wells, mesoscopic, and nanoscale system: Quantum dots
% 78.67.Hc Optical properties of low-dimensional structures: Quantum dots
% 73.22.-f Electronic structure of nanoscale materials: clusters,
%          nanoparticlehell, nanotubes, and nanocrystals
% 71.15.-m computational methodology use 71.15.-m
%nanoparticles, nanotubes, and nanocrystals

\begin{abstract}
We study the evolution of entanglement, quantum correlation and
classical correlation for the one dimensional XY model in external
transverse magnetic field. The system is initialized in the full
polarized state along the z axis, after annealing, different sites
will become entangled. We study the three kinds of correlation for
both the nearest and the next-nearest neighbor sites. We find that
for large anisotropy parameter the quantum phase transition can be
indicated by the dynamics of classical correlation between the
nearest neighbor sites. We find that the dynamics of entanglement
for both the nearest and next-nearest neighbor sites show
significantly different behaviors with different values of magnetic
field. We also find that the evolution of quantum correlation and
classical correlation of the nearest neighbor sites are obviously
different from those of the next-nearest neighbor sites.

\end{abstract}

\pacs{03.65.Ud, 75.10.pq, 87.16.dj}

\maketitle %\nopagebreak

As an ideal model used to describing phase transition on the early
days in statistical mechanics, the XY model has been studied
intensely for almost one century. Although it is so simple that it
can be calculated analytically, it catches the most important
elements of some physical system, such as materials
$\textmd{Cs}(\textmd{H}_{1-x}\textmd{D}_x)_2\textmd{PO}_4$,
$\textmd{PbH}_{1-x}\textmd{D}_x\textmd{PO}_4$ \cite{OD00}, etc. So
it could be used to depict the actual phase transition very well. It
also became the test bed equipped to describe more practice physical
system, and to examine the performance of different approximation
methods \cite{E70}. With the birth of quantum information theory, it
was found that the XY model could be used as a quantum information
channel. So it became the focus of theoretical and experimental
studying again \cite{Do07}, and many quantum information process has
been proposed with the XY model \cite{C04,A04,W07}.

There were many works on quantum information process in the XY and
the Ising model, such as the transfer of an unknown quantum state
along the chain \cite{C04,A04,B06}. As an important resource in
quantum information process, it was found that entanglement could be
used as an quantity to indicate quantum phase transition
\cite{AO02,O02}. Recently, it has been reported that dynamics of the
nearest-neighbor entanglement may be used as an indicator of quantum
phase transition in the transverse Ising model \cite{Zhe10}. In
fact, quantum entanglement is only a special part of the quantum
correlation \cite{H02,Ve03,J09,Ka10}, there is other quantum
correlation that is not caught by the quantum entanglement, and it
is found to be important in quantum information task
\cite{Me00,L08,Da08}. More correlation indicates more information,
so it is not trivial to study the evolution of the quantum
correlation.

In this paper, we study the entanglement, quantum correlation and
classical correlation dynamics of both the nearest and next-nearest
neighbor sites in the one dimensional transverse XY model with the
initial state been the full polarized state along the z axis. For
different values of anisotropy parameter, we find that the
entanglement takes on distinctive behaviors with different values of
magnetic field. We find that for large value of anisotropy parameter
the first maximum of classical correlation between the nearest
neighbor sites peaks around the critical point. That may be an
indicator of quantum phase transition. Further more, we find that
behaviors of the three kinds of correlation for the nearest neighbor
sites are obviously different from those for the next-nearest
neighbor sites.

The Hamiltonian of the one dimensional XY model in transverse
magnetic field reads as
\begin{equation}
H=-\lambda\sum_{i=1}^N[(1+\gamma)S_i^xS_{i+1}^x+(1-\gamma)S_i^yS_{i+1}^y]-\sum_{i=1}^NS_i^z,
\end{equation}
where $S_i^{\alpha}$ $(\alpha = x, y, z)$ are the spin $1/2$
matrixes at site $i$, and $N$ is the number of sites.  We assume
periodic boundary conditions $S_i^{\alpha}=S_{N+i}^{\alpha}$, and
the anisotropy parameter $\gamma$ divides the whole parameter space
into two classes, for $\gamma=0$, it is the XX model, and
$0<\gamma\leq1$ belongs to the Ising universality class. Here we
will only concentrate on the thermodynamic limit $N=\infty$, at
which the Hamiltonian system undergoes a quantum phase transition at
$\lambda_c=1$.

This model could be solved analytically by mapping the spin half
system into the spinless fermion system via the Jordan-Wigner
transformation \cite{E70}
\begin{eqnarray}
S_j^x&=&\frac12(b^{\dagger}_jexp(i\pi\sum_{k=1}b^{\dagger}_kb_k)+exp(-i\pi\sum_{k=1}b^{\dagger}_kb_k)b_j),\nonumber\\
S_j^y&=&\frac{1}{2i}(b^{\dagger}_jexp(i\pi\sum_{k=1}b^{\dagger}_kb_k)-exp(-i\pi\sum_{k=1}b^{\dagger}_kb_k)b_j),\nonumber\\
S_j^z&=&b^{\dagger}_jb_j-\frac12,
\end{eqnarray}
where $b_j^{\dagger}$ ($b_j$) is the creation (annihilation)
operator of the spinless fermion. After application of the Fourier
and Bogoliubov transformation
$\eta=\frac{1}{\sqrt{N}}\sum_le^{ikl}(\alpha_kb_l+i\beta_kb^{\dagger}_l)$,
with parameters $\alpha_k=\frac{\Lambda_k-(1+\lambda
\textsf{cos}k)}{\sqrt{2[\Lambda_k^2-(1+\lambda
\textsf{cos}k)\Lambda_k]}}$, $\beta_k=\frac{\gamma\lambda
\textsf{sin}k}{\sqrt{2[\Lambda_k^2-(1+\lambda
\textsf{cos}k)\Lambda_k]}}$, the eventually Hamiltonian can be
diagonalized as
\begin{equation}
H=\sum_k\Lambda_k(\eta^{\dagger}_k\eta_k-\frac12),
\end{equation}
with the eigenenergy spectrum $\Lambda_k=\sqrt{(1+\lambda
\textsf{cos}k)^2+\lambda^2\gamma^2 \textsf{sin}^2k}$.

The time evolution of the spin system in Eq. (1), can be given by
the evolution of the spinless fermion operator $b_j(t)$ \cite{La04}
\begin{equation}
b_j(t)=\frac1N\sum_{k,l}[\textsf{cos}k(l-j)c_k(t)b_l-2i\textsf{sin}k(l-j)d_k(t)b^{\dagger}_l],
\end{equation}
with time dependent coefficients
$c_k(t)=e^{i\Lambda_kt}-2i\beta^2_k\textsf{sin}\Lambda_kt$,
$d_k(t)=\alpha_k\beta_k\textsf{sin}\Lambda_kt$.

We choose the initial state that all spins are polarized along the
$z$ axis, $\rho(0)=|N\rangle\langle N|$, with
$|N\rangle=|\uparrow\rangle_1...|\uparrow\rangle_N$, which is ground
state of Eq. (1) when a strong magnetic field is applied along the
$z$ axis. Hence, there is no quantum correlation and classical
correlation among any partition of the system. Due to the
translational symmetry of the system, we choose site 1 as our first
site when we study the correlation between different sites. In order
to obtain the evolution of entanglement, quantum correlation and
classical correlation, we need to get the reduced density matrix
first. So we will have to calculate two-body spin correlation
functions, it is equal to calculate the fermion correlation function
$\langle b_1...\rangle_t=\langle b_1(t)...\rangle_0$. From Wick
theorem, we can learn that the correlation function will be nonzero
for only even number of creation and annihilation operators.

For the first two spins the reduced density matrix in the basis of
$|\uparrow\uparrow\rangle, |\uparrow\downarrow\rangle,
|\downarrow\uparrow\rangle, |\downarrow\downarrow\rangle,$ is given
by
\begin{eqnarray}
\rho^{12}(t)=\begin{pmatrix} \rho^{12}_{11} &0&0&\rho^{12}_{14}\\
0&\rho^{12}_{22}&\rho^{12}_{23}&0\\
0&\rho^{12}_{32}&\rho^{12}_{33}&0\\
\rho^{12}_{41} &0&0&\rho^{12}_{44}
\end{pmatrix},
\end{eqnarray}
where \cite{Zhe10}
\begin{eqnarray}
\rho^{12}_{11}&=&\langle
b_1^{\dagger}(t)b_1(t)b_2^{\dagger}(t)b_2(t)\rangle_0=\langle
b_1^{\dagger}(t)b_1(t)\rangle_0\langle
b_2^{\dagger}(t)b_2(t)\rangle_0\nonumber\\
&-&\langle b_1^{\dagger}(t)b_2^{\dagger}(t)\rangle_0\langle
b_1(t)b_2(t)\rangle_0+\langle
b_1^{\dagger}(t)b_2(t)\rangle_0\nonumber\\
&\times&\langle
b_1(t)b_2^{\dagger}(t)\rangle_0,\nonumber\\
\rho^{12}_{22}&=&\rho^{12}_{33}=\langle
b_1^{\dagger}(t)b_1(t)b_2(t)b_2^{\dagger}(t)\rangle_0=\langle
b_1^{\dagger}(t)b_1(t)\rangle_0-\rho^{12}_{11},\nonumber\\
\rho^{12}_{44}&=&\langle
b_1(t)b_1^{\dagger}(t)b_2(t)b_1^{\dagger}(t)\rangle_0=1-\rho^{12}_{11}-2\rho^{12}_{22},\nonumber\\
\rho^{12}_{41}&=&\rho^{12*}_{14}=\langle
b_1^{\dagger}(t)b_2^{\dagger}(t)\rangle_0,\nonumber\\
\rho^{12}_{23}&=&\rho^{12*}_{32}=-\langle
b_1(t)b_2^{\dagger}(t)\rangle_0.
\end{eqnarray}

At the thermodynamic limit $N=\infty$, we get the following
analytical formulations,
\begin{eqnarray}
\langle
b_1^{\dagger}(t)b_1(t)\rangle_0&=&\frac{1}{2\pi}\int_{-\pi}^{\pi}(1-4\alpha_k^2\beta_k^2\textsf{sin}^2\Lambda_kt)dk,\nonumber\\
\langle
b_1^{\dagger}(t)b_2(t)\rangle_0&=&\frac{1}{2\pi}\int_{-\pi}^{\pi}\textsf{cos}k(1-4\alpha_k^2\beta_k^2\textsf{sin}^2\Lambda_kt)dk,\nonumber\\
\langle
b_1(t)b_2(t)\rangle_0&=&\frac{1}{2\pi}\int_{-\pi}^{\pi}\textsf{sin}k\alpha_k\beta_k(2\textsf{sin}^2\Lambda_kt(1-2\beta_k^2)\nonumber\\
&-&i\textsf{sin}2\Lambda_kt)dk.
\end{eqnarray}

\begin{figure}
\includegraphics[width=3in]{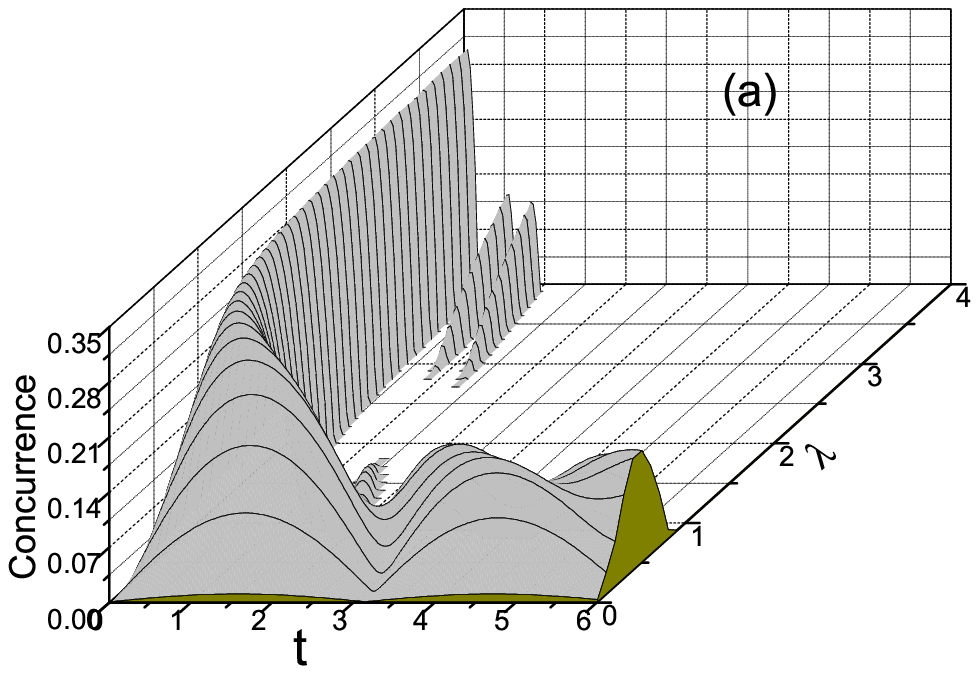}
\includegraphics[width=3in]{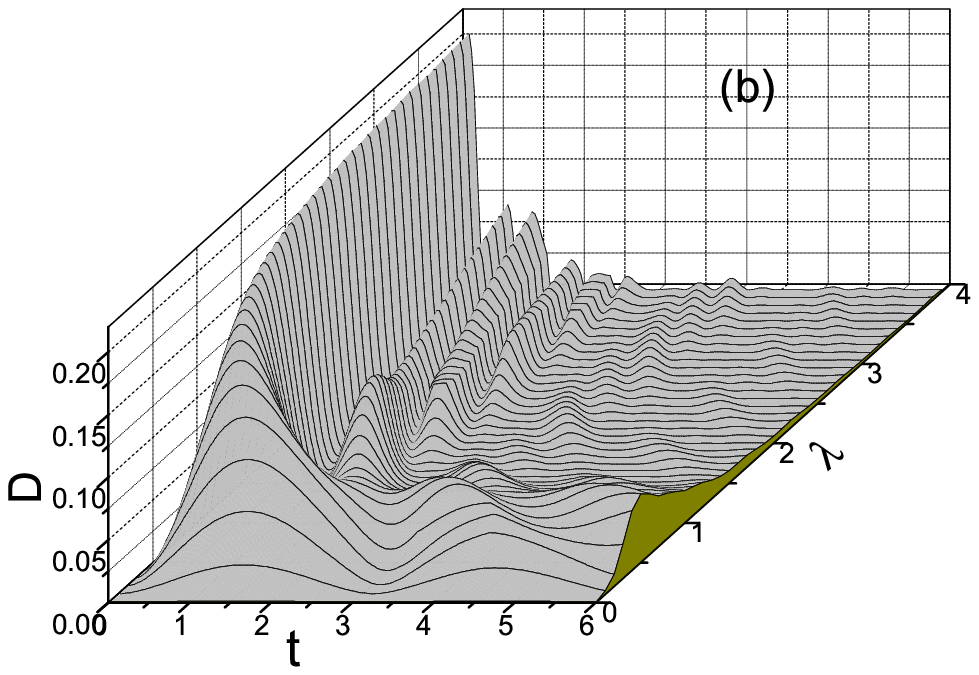}
\includegraphics[width=3in]{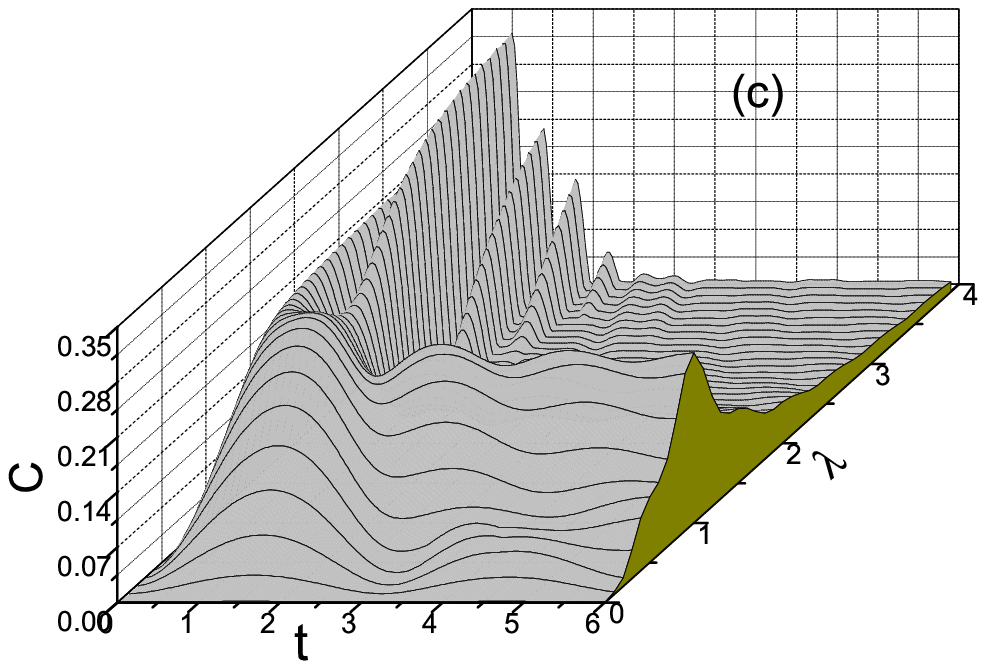}
\caption{Evolution of (a) entanglement, (b) quantum correlation and
(c) classical correlation for the nearest sites, with anisotropy
parameter $\gamma=1.0$.}
\end{figure}

We use concurrence \cite{Wo98} to measure the bipartite
entanglement. When the density matrix is the "X" type, it is given
by \cite{Yu07}
$2\textsf{max}[0,|\rho^{12}_{23}|-\sqrt{\rho^{12}_{11}\rho^{12}_{44}},
|\rho^{12}_{14}|-\sqrt{\rho^{12}_{22}\rho^{12}_{33}}]$. In order to
obtain the dynamics of quantum correlation and classical
correlation, we choose quantum discord \cite{H02} to measure the
quantum correlation (it equals to the quantum correlation \cite{H01}
in the case of two qubits \cite{J09,H01,Ge10}). It is given by the
discrepancy of two kinds of quantum mutual information derived from
their classical counterparts,
$D(\rho_{12})=I(\rho_{12})-C(\rho_{12})$, where
$C(\rho_{12})=\textsf{max}_{\{\Pi_j^2\}}(S(\rho_1)-\sum_jq_jS(\rho_1^j))$.
$\{\Pi_j^2\}$ are the complete projective measurement on partition
2, and $q_j=\textsf{tr}_{1,2}(\Pi_j^2\rho_{12})$.
$\rho_1^j=(\Pi_j^2\rho_{12}\Pi_j^2)/q_j$ is the state of partition 1
after partition 2 recording the outcome $j$.

\begin{figure}
\includegraphics[width=3in]{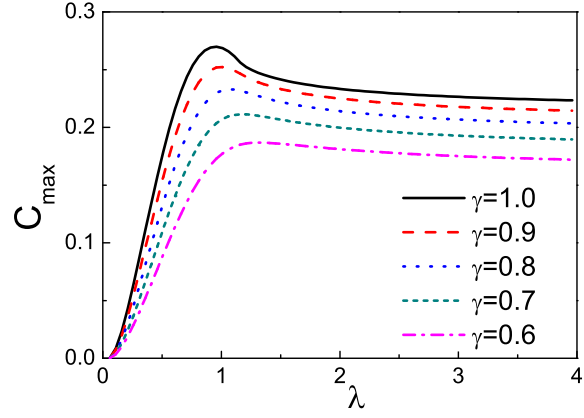}
\caption{(Color online) The first local maximum of classical
correlation for different values of anisotropy parameter. The solid
black, dash red, dot blue, short dash dark cyan line denote
$\gamma=1.0, 0.9, 0.8, 0.7$ respectively.}
\end{figure}

\begin{figure}
\includegraphics[width=3in]{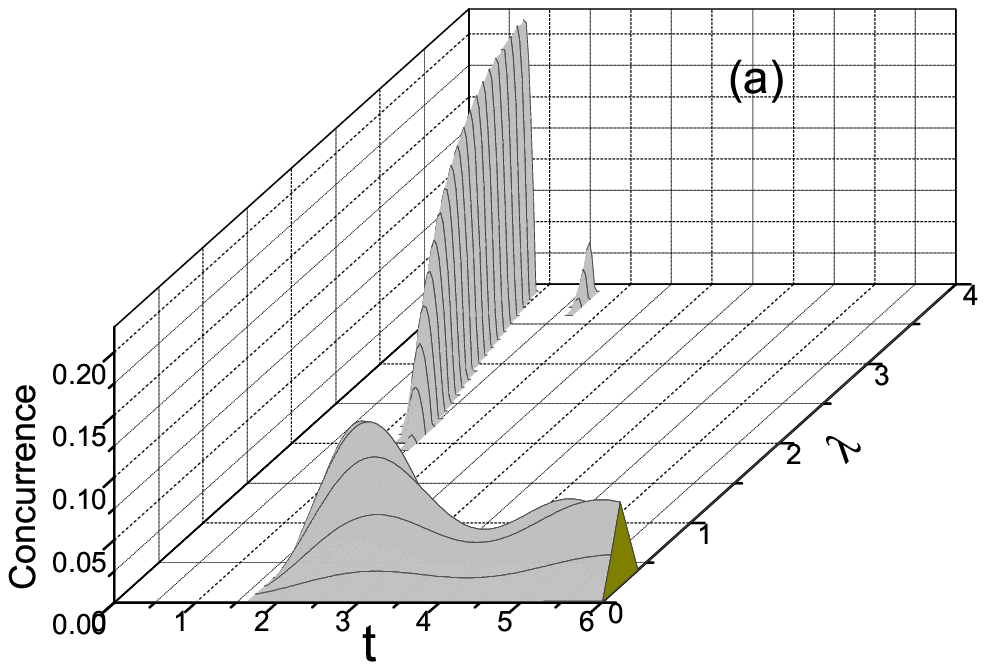}
\includegraphics[width=3in]{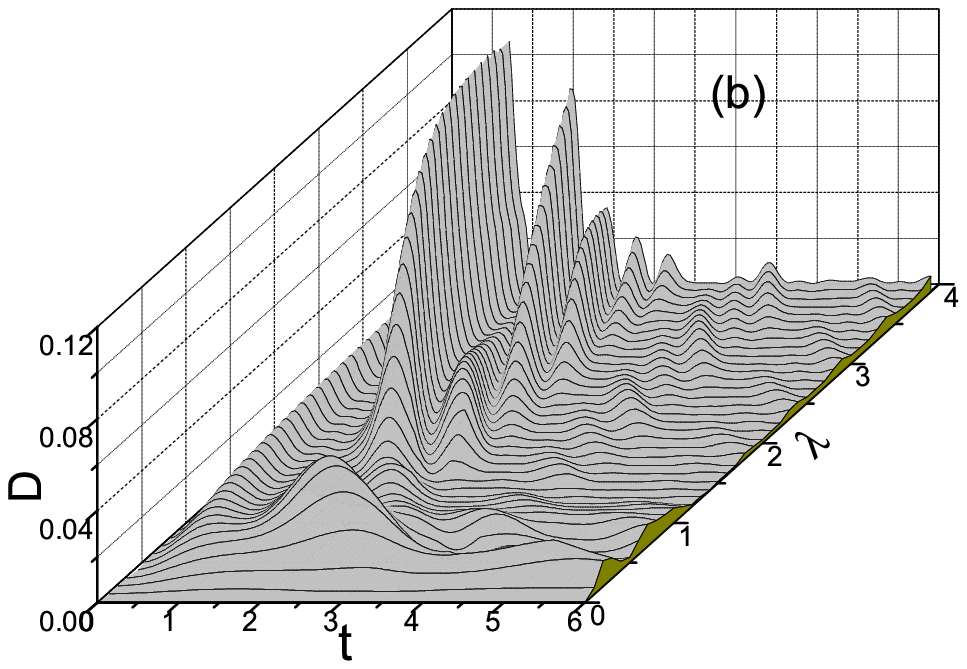}
\includegraphics[width=3in]{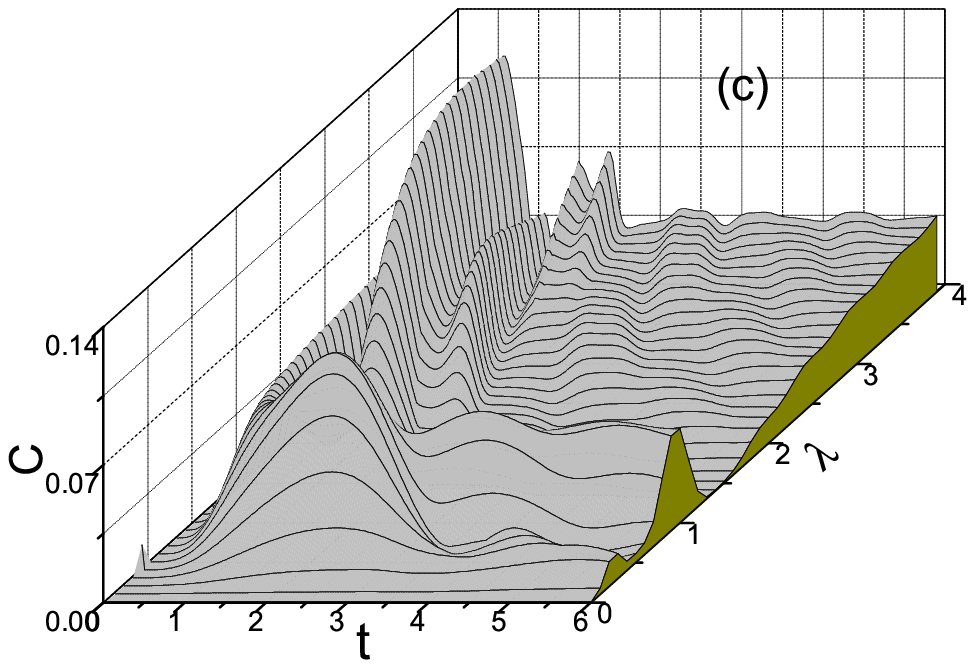}
\caption{The evolution of (a) entanglement, (b) quantum correlation
and (c) classical correlation for the next-nearest neighbor sites
with anisotropy parameter $\gamma=1.0$.}
\end{figure}

The initial state of the spin 1 and spin 2 reads
\begin{eqnarray}
\rho_{12}=\begin{pmatrix} 1&0&0&0\\
0&0&0&0\\
0&0&0&0\\
0&0&0&0\\
\end{pmatrix},
\end{eqnarray}
so there are neither classical correlation nor quantum correlation
between 1 and 2. As the magnetic field quenched ($\lambda $ changes
suddenly from zero to a finite value), the initial state will evolve
under the XY exchange interaction. In Figure 1(a), (b) and (C),
evolution of entanglement, classical correlation and quantum
correlation have been shown with the anisotropy parameter
$\gamma=1.0$. The behavior of entanglement is complex. There is a
definite boundary around $\lambda_b\sim0.86$, when
$\lambda\le\lambda_b$ entanglement is present in a long time
interval, but when $\lambda\ge\lambda_b$ entanglement disappears
fast and absents for a longtime. However, the quantum correlation is
nonzero almost all the time. So it is clear that it is due to the
dissonance \cite{Ka10}, another kind of quantum correlation, the
quantum correlation could be nonzero without entanglement. From the
evolution of both entanglement and quantum correlation we could see
that when the magnetic field is large ($\frac{1}{\lambda}$), there
will be entanglement all the time. While for a little magnetic
field, the dissonance, will be the dominant component of quantum
correlation and evolves with time. The classical correlation behaves
much the same like the quantum correlation, as the magnetic field
decreases classical correlation oscillates much faster, it can be
understood as the magnetic field decreases the flipping interaction
dominates over the on site potential energy. It can be seen that for
all kinds of correlation, they disappear gradually as the magnetic
field dying out. This is due to the fact the exchange interaction
tend to dispose the nearest sites in the sparable state
$(|\uparrow\rangle_1+|\downarrow\rangle_1)\otimes(|\uparrow\rangle_2-|\downarrow\rangle_2)$
or
$(|\uparrow\rangle_1-|\downarrow\rangle_1)\otimes(|\uparrow\rangle_2+|\downarrow\rangle_2)$.

For different values of the magnetic field, entanglement, quantum
correlation and classical correlation evolve with time, and they
will obtain their first local maximum with different values.
Particularly, we find that for the classical correlation, its first
local maximum $C_{max}$ has a maximum around the critical point
$\lambda_c=1$. This may be seen as an indicator of the quantum phase
transition at the critical point. In fact we find that when the
anisotropy parameter is big enough, $\gamma\geq0.7$, $C_{max}$
always has a maximum around the critical point, as shown in Fig. 2.
But as the anisotropy parameter decreases, the maximum gets away
from the critical point.

For the next-nearest neighbor sites, we only need to focus on sites
1 and 3. The reduced density matrix takes on the same shape as that
of sites 1 and 2, with 3 instead of 2 for the diagonal elements, and
the non-diagonal elements are given by
\begin{eqnarray}
\rho^{13}_{23}&=&\rho^{13*}_{32}=-\langle
b_1(t)b_3^{\dagger}(t)\rangle_0+2(\langle
b_1(t)b_3^{\dagger}(t)\rangle_0\langle
b_1^{\dagger}(t)b_1(t)\rangle_0\nonumber\\
&+&\langle b_1(t)b_2^{\dagger}(t)\rangle_0+|\langle
b_1(t)b_2(t)\rangle_0|^2),\nonumber\\
\rho^{13}_{14}&=&\rho^{13*}_{41}=-\langle
b_1(t)b_3(t)\rangle_0+2(\langle b_1(t)b_3(t)\rangle_0\langle
b_1^{\dagger}(t)b_1(t)\rangle_0\nonumber\\
&+&2Re(\langle b_1(t)b_2^{\dagger}(t)\rangle_0)\langle
b_1(t)b_2(t)\rangle_0,
\end{eqnarray}
where the transverse symmetry has been used, and $Re$ means the real
part of the number. The correspondent correlation functions are
\begin{eqnarray}
\langle
b_1^{\dagger}(t)b_3(t)\rangle_0&=&\frac{1}{2\pi}\int_{-\pi}^{\pi}\textsf{cos}2k(1-4\alpha_k^2\beta_k^2\textsf{sin}^2\Lambda_kt)dk,\nonumber\\
\langle
b_1(t)b_3(t)\rangle_0&=&\frac{1}{2\pi}\int_{-\pi}^{\pi}\textsf{sin}2k\alpha_k\beta_k(2\textsf{sin}^2\Lambda_kt(1-2\beta_k^2)\nonumber\\
& &-i\textsf{sin}2\Lambda_kt)dk.
\end{eqnarray}

Here for sites 1 and 3 the initial state is also represented by Eq.
(7), the dynamics of entanglement is obviously different from that
of nearest neighbor sites, for large anisotropy parameter the
entanglement does not appear until after a finite time interval, as
shown in Fig. 3(a). This is a little counterintuitive, because
entanglement between sites 1 (2) and 2 (3) comes into being almost
instantly due to their direct interaction. Therefore, it is
reasonable for the entanglement between sites 1 and 3 increases from
zero once sites 1(2) and 2(3) are entangled. At the same time, there
is a small band of $\lambda$ depending on the anisotropy parameter,
in which no entanglement will be generated. What is more, for both
classical correlation and quantum correlation, far away from the
critical point, they become larger as the magnetic field diminishes,
as shown in Figures 3(b), 3(c). This is a little unbelievable at the
first glance, and is obviously different from the case of the
nearest neighbor sites. At this case we do not find that the
dynamics of entanglement, quantum discord and classical correlation
could be used as indicator of the quantum phase transition at the
critical point like the nearest neighbor sites.

In this paper, we have studied the dynamics of entanglement, quantum
correlation and classical correlation with one dimensional
transverse XY model for both the nearest and next-nearest neighbor
sites. We found that for the nearest neighbor sites, when the
magnetic field is strong, all three kinds of correlation are larger
than those with a weak field. While for the next-nearest neighbor
sites, the dynamics is obviously different. For the nearest neighbor
sites, two kinds of dynamics are shown with different values of the
magnetic field, while for the next-nearest one, there is a narrow
band of $\lambda$, in which there is no entanglement generated. What
is more, we found that for the nearest neighbor sites, the first
maximum $C_{max}$ of classical correlation peaks around the critical
point for large anisotropy parameter, it may be used as an indicator
of quantum phase transition.

This work was supported by the National Fundamental Research
Program, National Natural Science Foundation of China (Grant Nos.
60921091 and 10874162). We thank H. Wichterich for helpful comments
and correspondence.

\end{document}